\documentclass[aps,prd,superscriptaddress,twoside,showpacs,nofootinbib,10pt,floatfix,showpacs,twocolumn]{revtex4-1}

\usepackage{bm}
\usepackage{tikz}
\usepackage{multirow}
\usepackage{amsmath}
\usepackage{gensymb}
\usepackage{graphicx}

\newcommand{\RNum}[1]{\uppercase\expandafter{\romannumeral #1\relax}}

\begin{document}
\title{Study of $qqq\bar{q}Q$ pentaquark system in the Chiral Quark Model}
\author{Qi Zhang}
\affiliation{
Department of Physics, Nanjing Normal University, Nanjing 210023, PR China
}
\author{Xiao-Huang Hu}
\affiliation{
Department of Physics, Nanjing Normal University, Nanjing 210023, PR China
}
\author{Bing-Ran He}
\email[E-mail: ]{hebingran@njnu.edu.cn (Corresponding author)}
\affiliation{
Department of Physics, Nanjing Normal University, Nanjing 210023, PR China
}
\author{Jia-Lun Ping}
\email[E-mail: ]{jlping@njnu.edu.cn (Corresponding author)}
\affiliation{
Department of Physics, Nanjing Normal University, Nanjing 210023, PR China
}
\date{\today}

\begin{abstract}
With the discovery of some hidden-charm pentaquark resonances by the LHCb Collaboration, investigations of pentaquark states containing heavy quarks have aroused the interest of theorists. We study herein $qqq\bar{q}Q$ ($q = u$ or $d$, $Q=c$ or $b$) pentaquark system, in the framework of the chiral quark model. In consequence, some charmed and bottomed pentaquarks are considered to exist by five-body dynamical calculations. In the charm sector, $\Sigma_c\pi(IJ^P=0\frac{1}{2}^-)$ and $\Sigma_c^*\pi(IJ^P=0\frac{3}{2}^-)$ are possible candidates of $\Lambda_c(2595)$ and $\Lambda_c(2625)$, respectively. Besides, two high-spin states, $\Sigma_c^*\rho(IJ^P=0\frac{5}{2}^-)$ and $\Delta D^*(IJ^P=1\frac{5}{2}^-)$, are also found in the energy region of $3.2 \sim 3.3$ GeV.
In the bottom sector, $\Sigma_b\pi(IJ^P=0\frac{1}{2}^-)$, $\Sigma_b^*\pi(IJ^P=0\frac{3}{2}^-)$ could be candidates of $\Lambda_b(5912)$ and $\Lambda_b(5920)$, respectively. And $\Sigma_b^*\rho(IJ^P=0\frac{5}{2}^-)$ and $\Delta B^*(IJ^P=1\frac{5}{2}^-)$ are found in the energy region of $6.5 \sim 6.6$ GeV. $\Sigma_c^{(*)}\pi$ and $\Sigma_b^{(*)}\pi$ are expected as compact states, while $\Sigma_c^*\rho$, $\Sigma_b^*\rho$, $\Delta D^*$ and $\Delta B^*$ are expected as molecular states.
\end{abstract}

\maketitle

\section{\label{sec:level1}Introductin}
Although the conventional quark model has achieved great success in describing the properties of ground states in the baryon sector with simple three-quark ($qqq$) configuration, it encounters difficulties in explaining some excited baryons with one quark in an orbital angular excitation. Faced with this situation, theorists have made an attempt to study the baryon structure with modified physical pictures that go beyond the simple quenched 3-quark models. In fact the spatial excitation energy of excited baryons is already sufficient to pull a $q\bar{q}$ pair from the gluon field, the multiquark system formed by this mechanism is considered to be more favorable than the P-wave excitation in the classical three-quark picture~\cite{Zou:2009wp}.

So far, many theoretical explanations for the excited baryons have been proposed from the unquenched picture. In the light baryon sector, the mass order reverse problem of $N^*(1535)$ and $\Lambda^*(1405)$ can be easily understood by considering them to be dynamically generated meson-baryon states~\cite{Kaiser:1995eg,Oller:2000fj,Inoue:2001ip,Magas:2005vu,Huang:2007zza,Bruns:2010sv,Qin:2020gxr} or states with large pentaquark components~\cite{Helminen:2000jb,Liu:2005pm,Zou:2005mq}. In the heavy baryon sector, some charmed baryons discovered in recent years, lie near their S-wave baryon-meson thresholds, which indicates that they are probably molecular pentaquark states with the exception of three-quark baryons in an orbital excitation. For instance, $\Lambda_c(2595)$ and $\Lambda_c(2625)$ are suggested to be a dynamically generated state of $\Sigma_c\pi$ and $\Sigma_c^*\pi$ coupling with other possible higher channels~\cite{Lu:2014ina,Lu:2016gev,Nieves:2019nol}, while they are also suggested to be generated states dominant by $ND$ and $ND^*$~\cite{Hofmann:2005sw,GarciaRecio:2008dp,Haidenbauer:2010ch,Romanets:2012hm,Liang:2014kra}. And the $\Lambda_c(2940)$ has been broadly discussed as a molecular state composed of a nucleon and $D^*$ meson~\cite{Zhao:2016zhf,He:2010zq,Ortega:2012cx,Zhang:2012jk}, while an isovector state of $\Sigma_c(2800)$, whose energy is a little below the $ND$ threshold, is still treated as a dynamically generated state from baryon-meson interactions~\cite{Lutz:2005ip,JimenezTejero:2009vq,Dong:2010gu,Sakai:2020psu}.
For the bottomed baryons, $\Lambda_b(5912)$ and $\Lambda_b(5920)$ are suggested to be $\Sigma_b\pi$ and $\Sigma_b^*\pi$ ~\cite{Lu:2016gev}, while in Refs.~\cite{GarciaRecio:2012db,Liang:2014eba}, the dominant coupled channel is identified as $N\bar{B}$ and $N\bar{B}^*$.
Despite many theoretical explanations to these excited baryons, their structures and properties are still controversial, thus it requires more efforts to investigate these states. On the experimental side, the several $P_c$ states were observed in the $J/\psi p$ invariant mass distribution of the $\Lambda_b^0 \rightarrow J/\psi K^-p$ decay by the LHCb Collaboration~\cite{Aaij:2015tga,Aaij:2019vzc}, which triggers an upsurge in investigations of pentaquark states containing heavy quarks.

Based on the above facts, we do a dynamic calculation of molecular five-quark system, composed of $qqq\bar{q}Q$ ($q = u$ or $d$, $Q=c$ or $b$), in the framework of the chiral quark model (ChQM). 
The chiral quark model has been used to study the recently observed hidden-charm pentaquarks~\cite{Yang:2015bmv}, and the results are supported by many other theoretical studies. In order to strictly solve Schr\"{o}dinger equation, a powerful method of few-body system, the Gaussian expansion method (GEM)~\cite{Hiyama:2003cu}, is employed to deal with orbital wave functions of five-body system. In the present work, because we are interested in the lowest-lying states of five-quark system, only the possible quantum numbers with negative parity are taken into account. In addition, we select the molecular configuration as the starting point of the research because: (1), masses of those charmed baryons with negative parity are very close to the relevant baryon-meson thresholds and molecular configuration is more physically possible compared with other configurations; (2), the energy of hidden color channel is generally higher than that of color-singlet channel, and the molecular configuration is the only possible way constructed by color-singlet structure. Two spatial configurations, [$qqQ$][$\bar{q}q$] and [$qqq$][$\bar{q}Q$], with all possible color, spin, flavor structures are considered in our calculation. Through the systematical search for bound states or possible resonant states, our purpose is to investigate whether some charmed baryons could be explained from the perspective of pentaquark states. On the other hand, an extension to the bottom case is performed to look for more states that may be observed in future experiments.

The paper is organized as follows. In Section.~\ref{sec:level2}, a brief introduction of the chiral quark model and the wave functions of the $qqq\bar{q}Q$ are presented. The numerical results are discussed in Section.~\ref{sec:level3}. In the last Section, we give a brief summary of this work.

\section{\label{sec:level2}Theoretical Framework}
\subsection{The chiral quark model}
The calculation is done in the framework of ChQM, which has been successfully applied in describing hadron spectrum and hadron-hadron interaction~\cite{Obukhovsky:1990tx,Fernandez:1993hx,Yu:1995ag,Valcarce:1995dm}. The Hamiltonian for multiquark system takes the form,
\begin{widetext}
\begin{eqnarray}
    && H = \sum_{i=1}^n \left( m_i + \frac{\bm{p}_i^2}{2m_i} \right) - T_{cm} + \sum_{j>i=1}^n \left( V_{ij}^C + V_{ij}^G + V_{ij}^{\pi} + V_{ij}^K + V_{ij}^{\eta} + V_{ij}^{\sigma} \right), \\
    && V_{ij}^C = \left( \bm{\lambda}_{i}^c \cdot \bm{\lambda}_{j}^c \right) \left[ -a_c \left( 1 - e^{-\mu_c r_{ij}} \right) + \Delta \right], \\
    && V_{ij}^G = \frac{1}{4} \alpha_s \left( \bm{\lambda}_{i}^c \cdot \bm{\lambda}_{j}^c \right)
                  \left[ \frac{1}{r_{ij}} - \frac{1}{6m_im_j} \frac{e^{-\frac{r_{ij}}{r_0(\mu_{ij})}}}{r_{ij}r_0^2(\mu_{ij})}
                  \bm{\sigma}_i \cdot \bm{\sigma}_j \right], \quad r_0(\mu_{ij}) = \hat{r}_0 / \mu_{ij}, \\
    && V_{ij}^{\sigma} = -\frac{g_{ch}^2}{4\pi} \frac{\Lambda_{\sigma}^2}{\Lambda_{\sigma}^2 - m_{\sigma}^2} m_{\sigma}
                         \left[ Y(m_{\sigma}r_{ij}) - \frac{\Lambda_{\sigma}}{m_{\sigma}} Y(\Lambda_{\sigma}r_{ij}) \right], \quad Y(x) = e^{-x}/x, \\
    && V_{ij}^{\pi} = \frac{g_{ch}^2}{4\pi} \frac{m_{\pi}^2}{12m_im_j} \frac{\Lambda_{\pi}^2}{\Lambda_{\pi}^2 - m_{\pi}^2} m_{\pi}
                      \left[ Y(m_{\pi}r_{ij}) - \frac{\Lambda_{\pi}^3}{m_{\pi}^3} Y(\Lambda_{\pi}r_{ij}) \right] \bm{\sigma}_i\cdot\bm{\sigma}_j \sum_{a=1}^3 \lambda_{i}^a \lambda_{j}^a, \\
    && V_{ij}^{K} = \frac{g_{ch}^2}{4\pi} \frac{m_{K}^2}{12m_im_j} \frac{\Lambda_{K}^2}{\Lambda_{K}^2 - m_{K}^2} m_{K}
                    \left[ Y(m_{K}r_{ij}) - \frac{\Lambda_{K}^3}{m_{K}^3} Y(\Lambda_{K}r_{ij}) \right] \bm{\sigma}_i\cdot\bm{\sigma}_j
                    \sum_{a=4}^7 \lambda_{i}^a \lambda_{j}^a, \\
    && V_{ij}^{\eta} = \frac{g_{ch}^2}{4\pi} \frac{m_{\eta}^2}{12m_im_j} \frac{\Lambda_{\eta}^2}{\Lambda_{\eta}^2 - m_{\eta}^2} m_{\eta}
                       \left[ Y(m_{\eta}r_{ij}) - \frac{\Lambda_{\eta}^3}{m_{\eta}^3} Y(\Lambda_{\eta}r_{ij}) \right] \bm{\sigma}_i\cdot\bm{\sigma}_j
                       \left( \lambda_{i}^8\lambda_{j}^8 \cos{\theta_p} - \lambda_{i}^0\lambda_{j}^0 \sin{\theta_p} \right).
\end{eqnarray}
\end{widetext}
Where $m_i$ is the mass of constituent quarks and $T_{cm}$ is the center-of-mass kinetic energy. The two-body potential includes phenomenology confinement potential $V^C$, one-gluon-exchange potential $V^G$, Goldstone boson exchange interactions $V^{\pi}$, $V^{K}$, $V^{\eta}$, and scalar meson potential $V^{\sigma}$. Note herein that the meson exchanges are considered when the two quarks are light, while the color confinement and the one-gluon exchange are flavor blindness. $\bm{\lambda}$ and $\bm{\lambda}^c$ are $SU(3)$ Gell-Mann matrices of flavor and color respectively, and $\bm{\sigma}$ is the $SU(2)$ Pauli matrix. $r_0(\mu_{ij})$ is a regulator that depends on $\mu_{ij}$. The chiral coupling constant $g_{ch}$ is determined from the pion-nucleon coupling constant and $Y(x)$ is the standard Yukawa functions. The specific introduction of the model can be found in Refs.~\cite{Vijande:2004he,Tan:2019qwe,Tan:2019knr}. In order to avoid the baryon-meson threshold order reversal, some hadrons should be described well in this work. The two sets of model parameters listed in Table~\ref{tab:table1} are adjusted by fitting the hadrons that we need. And some mesons and baryons involved in the present work are calculated with these parameters as shown in Table~\ref{tab:table2}.
\begin{table}[h]
\caption{\label{tab:table1}Quark Model Parameters.}
\begin{ruledtabular}
\begin{tabular}{ c c c c }
    \multirow{3}*{Quark masses} & $m_q = m_{\bar{q}}(MeV)$ & 450 & 430 \\
    & $m_c(MeV)$ & 1750 & 1734 \\
    & $m_b(MeV)$ & 5100 & 5092 \\
    \hline
    \multirow{3}*{Confinement} & $a_c(MeV)$ & 80 & 74 \\
    & $\mu_c(fm^{-1})$ & 0.7 & 0.7 \\
    & $\Delta(MeV)$ & 46 & 36 \\
    \hline
    \multirow{4}*{OGE} & $\alpha_s^{qq}$ & 0.67 & 0.68 \\
    & $\alpha_s^{qc}$ & 0.61 & 0.63 \\
    & $\alpha_s^{qb}$ & 0.59 & 0.61 \\
    & $\hat{r}_0(MeV \cdot fm)$ & 28.17 & 28.17 \\
    \hline
    \multirow{8}*{Goldstone bosons} & $m_{\pi}(fm^{-1})$ & 0.7 & 0.7 \\
    & $m_{K}(fm^{-1})$ & 2.51 & 2.51 \\
    & $m_{\eta}(fm^{-1})$ & 2.77 & 2.77 \\
    & $m_{\sigma}(fm^{-1})$ & 3.42 & 3.42 \\
    & $\Lambda_{\pi} = \Lambda_{\sigma}(fm^{-1})$ & 4.2 & 4.2 \\
    & $\Lambda_{K} = \Lambda_{\eta}(fm^{-1})$ & 5.2 & 5.2 \\
    & $g_{ch}^2/(4\pi)$ & 0.54 & 0.54 \\
    & $\theta_p(\degree)$ & -15 & -15 \\
\end{tabular}
\end{ruledtabular}
\end{table}
\begin{table}[h]
\caption{\label{tab:table2}The masses of the mesons and baryons involved in the calculation (unit: MeV).}
\begin{ruledtabular}
\begin{tabular}{ c c c c c c c c }
    Meson & \RNum{1} & \RNum{2} & Exp.~\cite{Zyla:2020zbs} & Baryon & \RNum{1} & \RNum{2} & Exp.~\cite{Zyla:2020zbs} \\
    \hline
    $\pi$ & $156$ & $149$ & $140$ & $N$ & $915$ & $908$ & $939$ \\
    $\rho$ & $795$ & $799$ & $775$ & $\Delta$ & $1229$ & $1235$ & $1232$ \\
    $\eta$ & $660$ & $672$ & $548$ & $\Lambda_c$ & $2270$ & $2263$ & $2286$ \\
    $\omega$ & $773$ & $777$ & $783$ & $\Sigma_c$ & $2469$ & $2473$ & $2455$ \\
    $D$ & $1867$ & $1866$ & $1869$ & $\Sigma_c^*$ & $2495$ & $2499$ & $2518$ \\
    $D^*$ & $2031$ & $2034$ & $2007$ & $\Lambda_b$ & $5586$ & $5589$ & $5620$ \\
    $B$ & $5303$ & $5315$ & $5279$ & $\Sigma_b$ & $5808$ & $5821$ & $5811$ \\
    $B^*$ & $5350$ & $5362$ & $5325$ & $\Sigma_b^*$ & $5818$ & $5832$ & $5832$ \\
\end{tabular}
\end{ruledtabular}
\end{table}

\subsection{The wave function}
\begin{figure}[h]
    \includegraphics[scale=1.0]{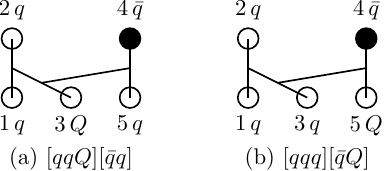}
    \caption{\label{fig:epsart}The molecular structures of $qqq\bar{q}Q$ system. The hollow circles stand for quark and the black circle stands for antiquark.}
\end{figure}
As shown in Fig.~\ref{fig:epsart}, there are two types of spatial configurations for $qqq\bar{q}Q$ system, one is the heavy quark in the 3-quark cluster denoted as [$qqQ$][$\bar{q}q$], the other is the heavy quark in the 2-quark cluster denoted as [$qqq$][$\bar{q}Q$]. The orbital wave function of the five-quark system can be obtained by coupling the orbital wave function of each relative motion, and is written as follows,
\begin{equation}
    \psi_{LM_L} = [ [ [ \psi_{l_1}(\bm{r}_{12}) \psi_{l_2}(\bm{r}_{12,3}) ]_l
                  \psi_{l_3}(\bm{r}_{45}) ]_{l'} \psi_{l_4}(\bm{r}_{123,45}) ]_{LM_L},
\end{equation}
where
\begin{eqnarray}
    \bm{r}_{12} &=& \bm{r}_{1} - \bm{r}_{2}\,, \nonumber\\
    \bm{r}_{12,3} &=& \frac{m_1\bm{r}_{1}+m_2\bm{r}_{2}}{m_1+m_2} - \bm{r}_{3}\,, \nonumber\\
    \bm{r}_{45} &=& \bm{r}_{4} - \bm{r}_{5}\,, \nonumber\\
    \bm{r}_{123,45} &=& \frac{m_1\bm{r}_{1} + m_2\bm{r}_{2} + m_3\bm{r}_{3}} {m_1+m_2+m_3} - \frac{m_4\bm{r}_{4}+m_5\bm{r}_{5}}{m_4+m_5}\,.\,\,
\end{eqnarray}
Here $\psi_{l_1}(\bm{r}_{12}), \psi_{l_2}(\bm{r}_{12,3}), \psi_{l_3}(\bm{r}_{45})$ are the orbital wave functions of relative motions in each cluster respectively, and $\psi_{l_4}(\bm{r}_{123,45})$ is the orbital wave function between two clusters, square brackets represent the angular momentum coupling. It is too difficult to obtain the orbital wave function by solving the Schr\"{o}dinger equation directly, so in our present work, GEM is employed to deal with orbital wave functions of five-body system. In GEM, the orbital wave function is written as the radial part and spherical harmonics, and the radial wave function is expanded by gaussians,
\begin{eqnarray}
    && \psi_{lm}(\bm{r}) = \sum_{n=1}^{n_{max}} c_n \psi_{nlm}^G(\bm{r}), \\
    && \psi_{nlm}^G(\bm{r}) = N_{nl} r^l e^{-\nu_{n}r^2} Y_{lm}(\hat{\bm{r}}), \\
    && N_{nl} = \left( \frac{2^{l+2}(2\nu_n)^{l+\frac{3}{2}}}{\sqrt{\pi}(2l+1)!!} \right)^{\frac{1}{2}}, \\
    && \nu_n = \frac{1}{r_n^2}, r_n = r_{min}a^{n-1}, a = \left( \frac{r_{max}}{r_{min}} \right)^{\frac{1}{n_{max}-1}}.
\end{eqnarray}
Where $c_n$ are expansion coefficients determined by solving the Schr\"{o}dinger equation; $N_{nl}$ are normalization constants; The Gaussian size $r_n$ is taken as the geometric progression to significantly reduce the dimension of matrix and simplify the calculation. For the present study, $n_{max}=7$ leads to a convergence result. Both the ground state energy and the radial excitation energy can be obtained by diagonalizing the Hamiltonian matrix.

When the angular momenta are not zero, the calculation of Hamiltonian matrix elements is rather complicated. It is convenient to employ the method of infinitesimally shifted Gaussians~\cite{Hiyama:2003cu},
\begin{eqnarray}
\psi_{nlm}^G(\bm{r}) && = N_{nl} r^l e^{-\nu_{n}r^2} Y_{lm}(\hat{\bm{r}}) \nonumber \\
                     && = N_{nl} \lim_{\epsilon \to 0} \frac{1}{(\nu_n \epsilon)^l} \sum_{k=1}^{k_{max}} C_{lm,k}
                                      e^{-\nu_n(\bm{r}-\epsilon\bm{D}_{lm,k})^2}.
\label{ISG}
\end{eqnarray}
the coefficients $C_{lm,k}$ and shift-direction vector $\bm{D}_{lm,k}$ are determined from the particular spherical harmonics. By absorbing the spherical harmonics into the shifted gaussians, complicated angular momentum algebra can be avoided.

The total spin wave function $\chi_{S,M_S}^{\sigma}(5)$ are obtained by coupling the spin wave functions of each cluster using Clebsch-Gordan Coefficients. Since the magnetic spin quantum number $M_S$ does not affect the energy of system, it is set to the maximum value here. There are five spin wave functions with $S = \frac{1}{2}$, four spin wave functions with $S = \frac{3}{2}$ and one spin wave function with $S = \frac{5}{2}$, respectively,
\begin{eqnarray}
    \chi_{\frac{1}{2},\frac{1}{2}}^{\sigma1}(5) = && \sqrt{\frac{1}{6}}\chi_{\frac{3}{2},-\frac{1}{2}}^{\sigma}(3)\chi_{1,1}^{\sigma}(2) -
                                                           \sqrt{\frac{1}{3}}\chi_{\frac{3}{2},\frac{1}{2}}^{\sigma}(3)\chi_{1,0}^{\sigma}(2)  \nonumber \\
                                                       &&  +\sqrt{\frac{1}{2}}\chi_{\frac{3}{2},\frac{3}{2}}^{\sigma}(3)\chi_{1,-1}^{\sigma}(2), \nonumber \\
    \chi_{\frac{1}{2},\frac{1}{2}}^{\sigma2}(5) = && \sqrt{\frac{1}{3}}\chi_{\frac{1}{2},\frac{1}{2}}^{\sigma1}(3)\chi_{1,0}^{\sigma}(2) -
                                                           \sqrt{\frac{2}{3}}\chi_{\frac{1}{2},-\frac{1}{2}}^{\sigma1}(3)\chi_{1,1}^{\sigma}(2), \nonumber \\
    \chi_{\frac{1}{2},\frac{1}{2}}^{\sigma3}(5) = && \sqrt{\frac{1}{3}}\chi_{\frac{1}{2},\frac{1}{2}}^{\sigma2}(3)\chi_{1,0}^{\sigma}(2) -
                                                           \sqrt{\frac{2}{3}}\chi_{\frac{1}{2},-\frac{1}{2}}^{\sigma2}(3)\chi_{1,1}^{\sigma}(2), \nonumber \\
    \chi_{\frac{1}{2},\frac{1}{2}}^{\sigma4}(5) = && \chi_{\frac{1}{2},\frac{1}{2}}^{\sigma1}(3)\chi_{0,0}^{\sigma}(2), \nonumber \\
    \chi_{\frac{1}{2},\frac{1}{2}}^{\sigma5}(5) = && \chi_{\frac{1}{2},\frac{1}{2}}^{\sigma2}(3)\chi_{0,0}^{\sigma}(2), \nonumber \\
    \chi_{\frac{3}{2},\frac{3}{2}}^{\sigma1}(5) = && \sqrt{\frac{3}{5}}\chi_{\frac{3}{2},\frac{3}{2}}^{\sigma}(3)\chi_{1,0}^{\sigma}(2) -
                                                           \sqrt{\frac{2}{5}}\chi_{\frac{3}{2},\frac{1}{2}}^{\sigma}(3)\chi_{1,1}^{\sigma}(2), \nonumber \\
    \chi_{\frac{3}{2},\frac{3}{2}}^{\sigma2}(5) = && \chi_{\frac{3}{2},\frac{3}{2}}^{\sigma}(3)\chi_{0,0}^{\sigma}(2), \nonumber \\
    \chi_{\frac{3}{2},\frac{3}{2}}^{\sigma3}(5) = && \chi_{\frac{1}{2},\frac{1}{2}}^{\sigma1}(3)\chi_{1,1}^{\sigma}(2), \nonumber \\
    \chi_{\frac{3}{2},\frac{3}{2}}^{\sigma4}(5) = && \chi_{\frac{1}{2},\frac{1}{2}}^{\sigma2}(3)\chi_{1,1}^{\sigma}(2), \nonumber \\
    \chi_{\frac{5}{2},\frac{5}{2}}^{\sigma1}(5) = && \chi_{\frac{3}{2},\frac{3}{2}}^{\sigma}(3)\chi_{1,1}^{\sigma}(2).
\end{eqnarray}
Where $\chi_{S,M_S}^{\sigma}(3)$ and $\chi_{S,M_S}^{\sigma}(2)$ are the spin wave functions of each sub-cluster,
\begin{eqnarray}
    && \chi_{\frac{3}{2},\frac{3}{2}}^{\sigma}(3) = \alpha\alpha\alpha, \nonumber \\
    && \chi_{\frac{3}{2},-\frac{3}{2}}^{\sigma}(3) = \beta\beta\beta, \nonumber \\
    && \chi_{\frac{3}{2},\frac{1}{2}}^{\sigma}(3) = \sqrt{\frac{1}{3}}
                                                        (\alpha\alpha\beta+\alpha\beta\alpha+\beta\alpha\alpha), \nonumber \\
    && \chi_{\frac{3}{2},-\frac{1}{2}}^{\sigma}(3) = \sqrt{\frac{1}{3}}
                                                           (\alpha\beta\beta+\beta\alpha\beta+\beta\beta\alpha), \nonumber \\
    && \chi_{\frac{1}{2},\frac{1}{2}}^{\sigma1}(3) = \sqrt{\frac{1}{6}}
                                                           (2\alpha\alpha\beta-\alpha\beta\alpha-\beta\alpha\alpha), \nonumber \\
    && \chi_{\frac{1}{2},\frac{1}{2}}^{\sigma2}(3) = \sqrt{\frac{1}{2}}
                                                           (\alpha\beta\alpha-\beta\alpha\alpha), \nonumber \\
    && \chi_{\frac{1}{2},-\frac{1}{2}}^{\sigma1}(3) = \sqrt{\frac{1}{6}}
                                                            (\alpha\beta\beta+\beta\alpha\beta-2\beta\beta\alpha), \nonumber \\
    && \chi_{\frac{1}{2},-\frac{1}{2}}^{\sigma2}(3) = \sqrt{\frac{1}{2}}
                                                            (\alpha\beta\beta-\beta\alpha\beta), \nonumber \\
    && \chi_{1,1}^{\sigma}(2) = \alpha\alpha, \nonumber \\
    && \chi_{1,-1}^{\sigma}(2) = \beta\beta, \nonumber \\
    && \chi_{1,0}^{\sigma}(2) = \sqrt{\frac{1}{2}}(\alpha\beta+\beta\alpha), \nonumber \\
    && \chi_{0,0}^{\sigma}(2) = \sqrt{\frac{1}{2}}(\alpha\beta-\beta\alpha).
\end{eqnarray}
Where $\alpha$ and $\beta$ denote spin-up and spin-down states, respectively.

The flavor wave function of five-quark system $\chi_{I,M_I}^{f}(5)$, can be obtained in the same way as the spin wave function. Here we take quark content $qqq\bar{q}c$ as an example,
\begin{eqnarray}
    \chi_{0,0}^{f1}(5) &=& \chi_{0,0}^{f}(3)\chi_{0,0}^{f}(2), \nonumber \\
    \chi_{0,0}^{f2}(5) &=& \sqrt{\frac{1}{2}} \chi_{\frac{1}{2},\frac{1}{2}}^{f1}(3)\chi_{\frac{1}{2},-\frac{1}{2}}^{f}(2)
                       - \sqrt{\frac{1}{2}} \chi_{\frac{1}{2},-\frac{1}{2}}^{f1}(3)\chi_{\frac{1}{2},\frac{1}{2}}^{f}(2), \nonumber \\
    \chi_{0,0}^{f3}(5) &=& \sqrt{\frac{1}{2}} \chi_{\frac{1}{2},\frac{1}{2}}^{f2}(3)\chi_{\frac{1}{2},-\frac{1}{2}}^{f}(2)
                       - \sqrt{\frac{1}{2}} \chi_{\frac{1}{2},-\frac{1}{2}}^{f2}(3)\chi_{\frac{1}{2},\frac{1}{2}}^{f}(2), \nonumber \\
    \chi_{0,0}^{f4}(5) &=& \sqrt{\frac{1}{3}} \chi_{1,1}^{f}(3)\chi_{1,-1}^{f}(2)
                       - \sqrt{\frac{1}{3}} \chi_{1,0}^{f}(3)\chi_{1,0}^{f}(2) \nonumber\\
                       &&+ \sqrt{\frac{1}{3}} \chi_{1,-1}^{f}(3)\chi_{1,1}^{f}(2), \nonumber\\
    \chi_{1,1}^{f1}(5) &=& \chi_{0,0}^{f}(3)\chi_{1,1}^{f}(2), \nonumber \\
    \chi_{1,1}^{f2}(5) &=& \chi_{1,1}^{f}(3)\chi_{0,0}^{f}(2), \nonumber \\
    \chi_{1,1}^{f3}(5) &=& \sqrt{\frac{1}{2}} \chi_{1,1}^{f}(3)\chi_{1,0}^{f}(2)
                       - \sqrt{\frac{1}{2}} \chi_{1,0}^{f}(3)\chi_{1,1}^{f}(2), \nonumber \\
    \chi_{1,1}^{f4}(5) &=& \chi_{\frac{1}{2},\frac{1}{2}}^{f1}(3)\chi_{\frac{1}{2},\frac{1}{2}}^{f}(2), \nonumber \\
    \chi_{1,1}^{f5}(5) &=& \chi_{\frac{1}{2},\frac{1}{2}}^{f2}(3)\chi_{\frac{1}{2},\frac{1}{2}}^{f}(2), \nonumber \\
    \chi_{1,1}^{f6}(5) &=& \sqrt{\frac{3}{4}} \chi_{\frac{3}{2},\frac{3}{2}}^{f}(3)\chi_{\frac{1}{2},-\frac{1}{2}}^{f}(2)
                       - \sqrt{\frac{1}{4}} \chi_{\frac{3}{2},\frac{1}{2}}^{f}(3)\chi_{\frac{1}{2},\frac{1}{2}}^{f}(2), \nonumber \\
    \chi_{2,2}^{f1}(5) &=& \chi_{1,1}^{f}(3)\chi_{1,1}^{f}(2), \nonumber \\
    \chi_{2,2}^{f2}(5) &=& \chi_{\frac{3}{2},\frac{3}{2}}^{f}(3)\chi_{\frac{1}{2},\frac{1}{2}}^{f}(2).
\end{eqnarray}
Where $\chi_{I,M_I}^{f}(3)$ and $\chi_{I,M_I}^{f}(2)$ are the flavor wave functions of each sub-cluster based on the SU(2) flavor symmetry,
\begin{eqnarray}
    && \chi_{0,0}^{f}(3) = \sqrt{\frac{1}{2}}(ud-du)c, \nonumber \\
    && \chi_{\frac{1}{2},\frac{1}{2}}^{f1}(3) = \sqrt{\frac{1}{6}}(2uud-udu-duu), \nonumber \\
    && \chi_{\frac{1}{2},\frac{1}{2}}^{f2}(3) = \sqrt{\frac{1}{2}}(udu-duu), \nonumber \\
    && \chi_{\frac{1}{2},-\frac{1}{2}}^{f1}(3) = \sqrt{\frac{1}{6}}(udd+dud-2ddu), \nonumber \\
    && \chi_{\frac{1}{2},-\frac{1}{2}}^{f2}(3) = \sqrt{\frac{1}{2}}(udd-dud), \nonumber \\
    && \chi_{1,1}^{f}(3) = uuc, \nonumber \\
    && \chi_{1,-1}^{f}(3) = ddc, \nonumber \\
    && \chi_{1,0}^{f}(3) = \sqrt{\frac{1}{2}}(ud+du)c, \nonumber \\
    && \chi_{\frac{3}{2},\frac{3}{2}}^{f}(3) = uuu, \nonumber \\
    && \chi_{\frac{3}{2},\frac{1}{2}}^{f}(3) = \sqrt{\frac{1}{3}}(uud+udu+duu), \nonumber \\
    && \chi_{0,0}^{f}(2) = \sqrt{\frac{1}{2}}(\bar{d}d+\bar{u}u), \nonumber \\
    && \chi_{\frac{1}{2},\frac{1}{2}}^{f}(2) = \bar{d}c, \nonumber
\end{eqnarray}
\begin{eqnarray}
    && \chi_{\frac{1}{2},-\frac{1}{2}}^{f}(2) = -\bar{u}c, \nonumber \\
    && \chi_{1,1}^{f}(2) = \bar{d}u, \nonumber \\
    && \chi_{1,-1}^{f}(2) = -\bar{u}d, \nonumber \\
    && \chi_{1,0}^{f}(2) = \sqrt{\frac{1}{2}}(\bar{d}d-\bar{u}u).
\end{eqnarray}

The physical pentaquark states must be colorless, but the way of reaching this condition can be acquired through the coupling of two colorless clusters or two colorful clusters. Therefore, there are two kinds of color structures, one is the color singlet-singlet $1 \bigotimes 1$ structure, and another is color octet-octet $8 \bigotimes 8$ structure. Color singlet wave function $\chi_1^c$, hidden color wave functions $\chi_2^c$ and $\chi_3^c$, can be found in Ref~\cite{Zhang:2020cdi}.

Finally, the total channel wave function of the five-quark system will be the product of orbit, spin, flavor and color components,
\begin{equation}
    \Psi_{JM_JIM_I}^{ijk} = \mathcal{A} \left[ \left[ \psi_L \chi_S^{\sigma i} \right]_{JM_J} \chi_I^{fj} \chi_k^c
                                       \right].
\label{eq_total_wave}
\end{equation}
The $i,j,k$ are the index of spin, flavor and color wave functions, respectively. Here, $\mathcal{A}$ is the antisymmetry operator, which ensures the antisymmetry of the total wave function when identical particles exchange. For the configuration of [$qqQ$][$\bar{q}q$] in Fig.~\ref{fig:epsart}, the antisymmetry operator is,
\begin{equation}
    \mathcal{A} = 1 - (15) - (25),
\end{equation}
and for the configuration of [$qqq$][$\bar{q}Q$],
\begin{equation}
    \mathcal{A} = 1 - (13) - (23).
\end{equation}
All possible physical channels for the $qqq\bar{q}c$ pentaquark system are listed in Table~\ref{tab:table3}. Superscript $S$ and $H$, denote color singlet channel and hidden color channel, respectively. The $i,j,k$ are the index in Eq.~\eqref{eq_total_wave}.

The eigen-energy is obtained by solving the Schr\"{o}dinger equation,
\begin{equation}
    H \Psi_{JM_JIM_I} = E \Psi_{JM_JIM_I}
\end{equation}
with the Rayleigh-Ritz variational principle.
\begin{table}[h]
    \caption{\label{tab:table3}Physical channels for $qqq\bar{q}c$ pentaquark system.}
    \resizebox{88mm}{99.7mm}{
    \begin{ruledtabular}
    \begin{tabular}{ c c c c c c }
        $IJ^P$ & $[i;j;k]$ & Channel & $IJ^P$ & $[i;j;k]$ & Channel \\
        \hline
        $0\frac{1}{2}^-$ & $[4;4;1]$ & $(\Sigma_c \pi)^S$ & $1\frac{1}{2}^-$ & $[5;1;1]$ & $(\Lambda_c \pi)^S$ \\
        {} & $[4,5;4;2,3]$ & $(\Sigma_c \pi)^H$ & {} & $[4,5;1;2,3]$ & $(\Lambda_c \pi)^H$ \\
        {} & $[5;1;1]$ & $(\Lambda_c \eta)^S$ & {} & $[4;3;1]$ & $(\Sigma_c \pi)^S$ \\
        {} & $[4,5;1;2,3]$ & $(\Lambda_c \eta)^H$ & {} & $[4,5;3;2,3]$ & $(\Sigma_c \pi)^H$ \\
        {} & $[3;1;1]$ & $(\Lambda_c \omega)^S$ & {} & $[4;2;1]$ & $(\Sigma_c \eta)^S$ \\
        {} & $[2,3;1;2,3]$ & $(\Lambda_c \omega)^H$ & {} & $[4,5;2;2,3]$ & $(\Sigma_c \eta)^H$ \\
        {} & $[2;4;1]$ & $(\Sigma_c \rho)^S$ & {} & $[3;1;1]$ & $(\Lambda_c \rho)^S$ \\
        {} & $[2,3;4;2,3]$ & $(\Sigma_c \rho)^H$ & {} & $[2,3;1;2,3]$ & $(\Lambda_c \rho)^H$ \\
        {} & $[1;4;1]$ & $(\Sigma_c^* \rho)^S$ & {} & $[2;3;1]$ & $(\Sigma_c \rho)^S$ \\
        {} & $[1;4;3]$ & $(\Sigma_c^* \rho)^H$ & {} & $[2,3;3;2,3]$ & $(\Sigma_c \rho)^H$ \\
        {} & $[4,5;2,3;1]$ & $(N D)^S$ & {} & $[2;2;1]$ & $(\Sigma_c \omega)^S$ \\
        {} & $[4,5;2,3;2,3]$ & $(N D)^H$ & {} & $[2,3;2;2,3]$ & $(\Sigma_c \omega)^H$ \\
        {} & $[2,3;2,3;1]$ & $(N D^*)^S$ & {} & $[1;3;1]$ & $(\Sigma_c^* \rho)^S$ \\
        {} & $[2,3;2,3;2,3]$ & $(N D^*)^H$ & {} & $[1;3;3]$ & $(\Sigma_c^* \rho)^H$ \\
        $0\frac{3}{2}^-$ & $[2;4;1]$ & $(\Sigma_c^* \pi)^S$ & {} & $[1;2;1]$ & $(\Sigma_c^* \omega)^S$ \\
        {} & $[2;4;3]$ & $(\Sigma_c^* \pi)^H$ & {} & $[1;2;3]$ & $(\Sigma_c^* \omega)^H$ \\
        {} & $[4;1;1]$ & $(\Lambda_c \omega)^S$ & {} & $[4,5;4,5;1]$ & $(N D)^S$ \\
        {} & $[3,4;1;2,3]$ & $(\Lambda_c \omega)^H$ & {} & $[4,5;4,5;2,3]$ & $(N D)^H$ \\
        {} & $[3;4;1]$ & $(\Sigma_c \rho)^S$ & {} & $[2,3;4,5;1]$ & $(N D^*)^S$ \\
        {} & $[3,4;4;2,3]$ & $(\Sigma_c \rho)^H$ & {} & $[2,3;4,5;2,3]$ & $(N D^*)^H$ \\
        {} & $[1;4;1]$ & $(\Sigma_c^* \rho)^S$ & {} & $[1;6;1]$ & $(\Delta D^*)^S$ \\
        {} & $[1;4;3]$ & $(\Sigma_c^* \rho)^H$ & {} & $[1;6;3]$ & $(\Delta D^*)^H$ \\
        {} & $[3,4;2,3;1]$ & $(N D^*)^S$ & $1\frac{3}{2}^-$ & $[2;3;1]$ & $(\Sigma_c^* \pi)^S$ \\
        {} & $[3,4;2,3;2,3]$ & $(N D^*)^H$ & {} & $[2;3;3]$ & $(\Sigma_c^* \pi)^H$ \\
        $0\frac{5}{2}^-$ & $[1;4;1]$ & $(\Sigma_c^* \rho)^S$ & {} & $[4;1;1]$ & $(\Lambda_c \rho)^S$ \\
        {} & $[1;4;3]$ & $(\Sigma_c^* \rho)^H$ & {} & $[3,4;1;2,3]$ & $(\Lambda_c \rho)^H$ \\
        $2\frac{1}{2}^-$ & $[4;1;1]$ & $(\Sigma_c \pi)^S$ & {} & $[2;2;1]$ & $(\Sigma_c^* \eta)^S$ \\
        {} & $[4,5;1;2,3]$ & $(\Sigma_c \pi)^H$ & {} & $[2;2;3]$ & $(\Sigma_c^* \eta)^H$ \\
        {} & $[2;1;1]$ & $(\Sigma_c \rho)^S$ & {} & $[3;3;1]$ & $(\Sigma_c \rho)^S$ \\
        {} & $[2,3;1;2,3]$ & $(\Sigma_c \rho)^H$ & {} & $[3,4;3;2,3]$ & $(\Sigma_c \rho)^H$ \\
        {} & $[1;1;1]$ & $(\Sigma_c^* \rho)^S$ & {} & $[3;2;1]$ & $(\Sigma_c \omega)^S$ \\
        {} & $[1;1;3]$ & $(\Sigma_c^* \rho)^H$ & {} & $[3,4;2;2,3]$ & $(\Sigma_c \omega)^H$ \\
        {} & $[1;2;1]$ & $(\Delta D^*)^S$ & {} & $[1;3;1]$ & $(\Sigma_c^* \rho)^S$ \\
        {} & $[1;2;3]$ & $(\Delta D^*)^H$ & {} & $[1;3;3]$ & $(\Sigma_c^* \rho)^H$ \\
        $2\frac{3}{2}^-$ & $[2;1;1]$ & $(\Sigma_c^* \pi)^S$ & {} & $[1;2;1]$ & $(\Sigma_c^* \omega)^S$ \\
        {} & $[2;1;3]$ & $(\Sigma_c^* \pi)^H$ & {} & $[1;2;3]$ & $(\Sigma_c^* \omega)^H$ \\
        {} & $[3;1;1]$ & $(\Sigma_c \rho)^S$ & {} & $[3,4;4,5;1]$ & $(N D^*)^S$ \\
        {} & $[3,4;1;2,3]$ & $(\Sigma_c \rho)^H$ & {} & $[3,4;4,5;2,3]$ & $(N D^*)^H$ \\
        {} & $[1;1;1]$ & $(\Sigma_c^* \rho)^S$ & {} & $[2;6;1]$ & $(\Delta D)^S$ \\
        {} & $[1;1;3]$ & $(\Sigma_c^* \rho)^H$ & {} & $[2;6;3]$ & $(\Delta D)^H$ \\
        {} & $[2;2;1]$ & $(\Delta D)^S$ & {} & $[1;6;1]$ & $(\Delta D^*)^S$ \\
        {} & $[2;2;3]$ & $(\Delta D)^H$ & {} & $[1;6;3]$ & $(\Delta D^*)^H$ \\
        {} & $[1;2;1]$ & $(\Delta D^*)^S$ & $1\frac{5}{2}^-$ & $[1;3;1]$ & $(\Sigma_c^* \rho)^S$ \\
        {} & $[1;2;3]$ & $(\Delta D^*)^H$ & {} & $[1;3;3]$ & $(\Sigma_c^* \rho)^H$ \\
        $2\frac{5}{2}^-$ & $[1;1;1]$ & $(\Sigma_c^* \rho)^S$ & {} & $[1;2;1]$ & $(\Sigma_c^* \omega)^S$ \\
        {} & $[1;1;3]$ & $(\Sigma_c^* \rho)^H$ & {} & $[1;2;3]$ & $(\Sigma_c^* \omega)^H$ \\
        {} & $[1;2;1]$ & $(\Delta D^*)^S$ & {} & $[1;6;1]$ & $(\Delta D^*)^S$ \\
        {} & $[1;2;3]$ & $(\Delta D^*)^H$ & {} & $[1;6;3]$ & $(\Delta D^*)^H$ \\
    \end{tabular}
    \end{ruledtabular}
    }
\end{table}

\section{\label{sec:level3}Numerical Results}
First, we study odd parity charmed and bottomed baryons that have been experimentally observed from three-quark system in an orbital excitation. The wave function is constructed in the same way as described above. Nevertheless, the orbital wave function of the three-quark system is written as 
$\psi_{lm_l} = [ \psi_{l_1}(\bm{r}_{12}) \psi_{l_2}(\bm{r}_{12,3}) ]_l$, the wave functions of two relative motions, $\psi_{l_1}(\bm{r}_{12})$ and $\psi_{l_2}(\bm{r}_{12,3})$, are expanded by infinitesimally shifted gaussian function according to Eq.~\eqref{ISG}. For the P-wave three-quark system, two low-lying excitation modes are considered in the calculation. One is the excitation in the first relative motion, denoted as $(l_1,l_2)=(1,0)$; the other is the excitation in the second relative motion, denoted as $(l_1,l_2)=(0,1)$. The energy of P-wave baryon is obtained by coupling the two excitation modes.

The low-lying orbital excited energies of $qqQ$ baryons with two sets of parameters are calculated and listed in Table~\ref{tab:Pwave}. The energies of two excitation modes and their coupling are given respectively, while for $\Sigma_c^*$ and $\Sigma_b^*$ states, only one excitation mode is allowed in order to fulfill the Pauli principle. When the orbital angular momentum is set to one, the spin can couple with it to give total quantum numbers $J^P=\frac{1}{2}^-$, $J^P=\frac{3}{2}^-$ or $J^P=\frac{5}{2}^-$, we obtain the same energy for these states because the spin-orbit interaction is not involved in the present model. So only states with $J^P=\frac{1}{2}^-$ are given. Considering small effect of spin-orbit interaction, it is enough to make a qualitative analysis. Although ground states of $qqQ$ baryons in Table~\ref{tab:table2} are described well in this work, the results in Table~\ref{tab:Pwave} show that the energies of P-wave excitation are about $130$ MeV lower than the experimental states with $J^P=\frac{1}{2}^-$. With regard to states with $J^P=\frac{3}{2}^-$, we also judge that there is a big deviation between the theoretical calculation and the experimental values even if the spin-orbit interaction is included. Therefore, we consider that it is inappropriate to explain these odd-parity excited states from the quenched three quark picture within our framework.
\begin{table}[h]
    \caption{\label{tab:Pwave}The low-lying orbital excited energies of $qqQ$ baryons (unit: MeV).}
    \begin{ruledtabular}
    \begin{tabular}{ c c c c c c }
        State & $IJ^P$ & $(l_1,l_2)$ & \RNum{1} & \RNum{2} & Exp.~\cite{Zyla:2020zbs} \\
        \hline
        $\Lambda_c$ & $0\frac{1}{2}^-$ & $(1,0)$ & $2667$ & $2665$ & {} \\
        {} & {} & $(0,1)$ & $2465$ & $2456$ & {} \\
        {} & {} & $coupling$ & $2465$ & $2456$ & $2592$ \\
        $\Sigma_c$ & $1\frac{1}{2}^-$ & $(1,0)$ & $2671$ & $2669$ & {} \\
        {} & {} & $(0,1)$ & $2629$ & $2628$ & {} \\
        {} & {} & $coupling$ & $2628$ & $2628$ & {} \\
        $\Sigma_c^*$ & $1\frac{1}{2}^-$ & $(0,1)$ & $2633$ & $2632$ & {} \\
        \hline
        $\Lambda_b$ & $0\frac{1}{2}^-$ & $(1,0)$ & $6006$ & $6012$ & {} \\
        {} & {} & $(0,1)$ & $5783$ & $5785$ & {} \\
        {} & {} & $coupling$ & $5783$ & $5785$ & $5912$ \\
        $\Sigma_b$ & $1\frac{1}{2}^-$ & $(1,0)$ & $6008$ & $6015$ & {} \\
        {} & {} & $(0,1)$ & $5954$ & $5963$ & {} \\
        {} & {} & $coupling$ & $5954$ & $5963$ & {} \\
        $\Sigma_b^*$ & $1\frac{1}{2}^-$ & $(0,1)$ & $5956$ & $5965$ & {} \\
    \end{tabular}
    \end{ruledtabular}
\end{table}

Next we turn to the pentaquark system. In the present work, because we are interested in the lowest-lying states of five-quark system, the angular momentum of each relative motion is set to zero. The calculation of energy spectrums is performed and the channel-coupling effect of different physical channels are considered. Moreover, the energy components are calculated to identify which terms in the Hamiltonian making the state to be bound, and the root-mean-square (RMS) distances between quark pairs are also calculated to unravel the structure of the state. We first investigate the $qqq\bar{q}c$ system with all possible negative-parity quantum numbers, and analyze the results in detail in the following.

$\bm{I(J^P) = 0(\frac{1}{2}^-):}$ The results of the single channel and channel coupling calculations are shown in Table~\ref{tab:table4}. Two spatial structures considered are clearly marked in the first column. The second column gives the physical channels involved in the present work. The third column shows the single channel eigen-energy by solving the Schr\"{o}dinger equation. The forth column refers to the theoretical value of noninteracting baryon-meson threshold. The values of binding energies $E_B = E - E_{th}(Theo.)$ are listed in the fifth column only if $E_B<0$. In order to reduce the theoretical error that appeared in the calculation of the mesons and baryons, we shift the energy of state to $E' = E_{th}(Exp.) + E_B$ in the sixth column, where $E_{th}(Exp.)$ is the experimental value of noninteracting baryon and meson. The last two rows are energies of the channel coupling calculation and the percentages of dominant channels. For the channel coupling calculation, the corrected energy is given as $E' = E + \sum_{i} p_i [E_{th}^i(Exp.) - E_{th}^i(Theo.)]$, where $p_i$ is the percentage of various physical channels. One can see in Table~\ref{tab:table4} that all physical channels are unbound except the lowest channel of $\Sigma_c\pi$ in the single channel calculation, which has the binding energy of $-13$ MeV. Therefore, a bound state with $\Sigma_c\pi$ configuration should be considered to exist in this quantum number sector. Nevertheless, $\Sigma_c$ can decay into $\Lambda_c$ and $\pi$ through strong interaction, so this bound state will appear as a resonance by experimental observation. After correction, the energy of $\Sigma_c\pi$ state is 2582 MeV, which is close to the $\Lambda_c(2595)$ with mass $M = 2592.25\pm0.28$ MeV. In fact, the channels of the system influence each other, thus it is necessary to take into account the channel coupling effect. For the channel coupling calculation, the lowest state $\Sigma_c\pi$ is pushed down about 20 MeV when coupling with other higher channels. After calculation of the composition, we find that the color singlet channel of $\Sigma_c\pi$ is the main component with the percentage of $90.6\%$, while the hidden color channel of $\Lambda_c\omega$ and the color singlet channel of $ND$ are the minor components, accounting for $6.2\%$ and $2.2\%$, respectively. Since the $\Lambda_c(2595)$ is located very close to the $\Sigma_c\pi$ threshold, this observation leads us naturally to consider a predominant baryon-meson structure of this lowest-lying odd parity charmed baryon. In our present calculation, we conclude that this state dominated by the $\Sigma_c\pi$ can be a candidate of the charmed baryon $\Lambda_c(2595)$.
\begin{table}[h]
    \caption{\label{tab:table4}The energies of the  $qqq\bar{q}c$ pentaquark system with quantum numbers $I(J^P) = 0(\frac{1}{2}^-)$ (unit: MeV).}
    \begin{ruledtabular}
    \begin{tabular}{ c c c c c c }
        Structure & Channel & $E$ & $E_{th}$(Theo.) & $E_B$ & $E'$ \\
        \hline
        $[qqc][\bar{q}q]$ & $(\Sigma_c \pi)^S$ & $2612$ & $2625$ & $-13$ & $2582$ \\
        {} & $(\Sigma_c \pi)^H$ & $3273$ & {} & {} & {} \\
        {} & $(\Lambda_c \eta)^S$ & $2931$ & $2930$ & $0$ & $2834$ \\
        {} & $(\Lambda_c \eta)^H$ & $3119$ & {} & {} & {} \\
        {} & $(\Lambda_c \omega)^S$ & $3044$ & $3043$ & $0$ & $3069$ \\
        {} & $(\Lambda_c \omega)^H$ & $3006$ & {} & {} & {} \\
        {} & $(\Sigma_c \rho)^S$ & $3264$ & $3264$ & $0$ & $3230$ \\
        {} & $(\Sigma_c \rho)^H$ & $3285$ & {} & {} & {} \\
        {} & $(\Sigma_c^* \rho)^S$ & $3290$ & $3290$ & $0$ & $3293$ \\
        {} & $(\Sigma_c^* \rho)^H$ & $3262$ & {} & {} & {} \\
        $[qqq][\bar{q}c]$ & $(N D)^S$ & $2782$ & $2782$ & $0$ & $2808$ \\
        {} & $(N D)^H$ & $3191$ & {} & {} & {} \\
        {} & $(N D^*)^S$ & $2947$ & $2946$ & $0$ & $2946$ \\
        {} & $(N D^*)^H$ & $3180$ & {} & {} & {} \\
        \hline
        $coupling$ & {} & $2591$ & $2625$ & $-34$ & $2566$ \\
        {} & \multicolumn{5}{c}{$(\Sigma_c\pi)^S$:$90.6\%$ ; $(\Lambda_c\omega)^H$:$6.2\%$ ; others:$3.2\%$} \\
    \end{tabular}
    \end{ruledtabular}
\end{table}

To identify which terms in the Hamiltonian making the state to be bound, the contributions from each term of Hamiltonian for the color singlet channel of $\Sigma_c\pi$, are given in Table~\ref{tab:table5}. $\Sigma_c + \pi$ represents the noninteracting $\Sigma_c$ baryon and $\pi$ meson. The difference between the pentaquark state and two free corresponding hadrons is denoted as $\Delta_E$, which stands for contributions of interaction between two clusters. Since identical light $q$ quarks can be exchanged in two clusters, kinetic energy of relative motion between two clusters provides major repulsion, with the contribution about $55.1$ MeV. However, considerable attraction from one-gluon-exchange and $\sigma$ meson exchange can cancel out the whole repulsion, and the rest of the attraction makes two hadrons to be bound. Besides, confinement potential and $\pi$ meson exchange provide a weak attraction between two clusters, while $\eta$ meson exchange provides slight repulsion. There is no contribution from the $K$ meson exchange because no $s$ quark involved in this system. Therefore, one-gluon-exchange and $\sigma$ meson exchange contribute to the binding of the $\Sigma_c\pi$ state.
\begin{table}[h]
    \caption{\label{tab:table5}The energy components of $\Sigma_c\pi$ state (unit: MeV).}
    \begin{ruledtabular}
    \begin{tabular}{ c c c c c }
            {} & rest mass & kinetic & $V^C$ & $V^G$ \\
            \hline
            $\Sigma_c\pi$ & $3550.0$ & $4136.9$ & $-233.9$ & $-4667.7$ \\
            $\Sigma_c + \pi$ & $3550.0$ & $4081.8$ & $-232.2$ & $-4635.0$ \\
            $\Delta_E$ & $0.0$ & $55.1$ & $-1.7$ & $-32.7$ \\
            \hline
            {} & $V^{\pi}$ & $V^{K}$ & $V^{\eta}$ & $V^{\sigma}$ \\
            \hline
            $\Sigma_c\pi$ & $-279.8$ & $0.0$ & $214.4$ & $-107.9$ \\
            $\Sigma_c + \pi$ & $-278.0$ & $0.0$ & $212.1$ & $-73.6$ \\
            $\Delta_E$ & $-1.8$ & $0.0$ & $2.3$ & $-34.3$ \\
    \end{tabular}
    \end{ruledtabular}
\end{table}

The spacial configurations of the states are determined by the dynamical calculation. The root-mean-square (RMS) distances between any two quarks for the system are calculated and shown in Table~\ref{tab:table6}. We do the channel coupling calculation for two spatial structures $[qqc][\bar{q}q]$ and $[qqq][\bar{q}c]$, respectively. From the results in Table~\ref{tab:table6} we can see that: (1), for $[qqc][\bar{q}q]$, the distances between any two quarks are around $1$ fm. Similar distances between any two quarks owe to action of the antisymmetry operator, which makes identical quarks exchange between two clusters. As we can't distinguish the identical quarks in two sub-clusters, the distances $r_{qq}$, $r_{q\bar{q}}$ and $r_{qc}$, is actually an average effect. The real distances among the sub-cluster are estimated to be less than $1$ fm, however, the distance between non-identical quarks, $r_{c\bar{q}}$, can be treated as the distance between two clusters; (2), for $[qqq][\bar{q}c]$, identical quarks are restricted in one cluster, the distances among the sub-cluster are less than $1$ fm, while the distances between two clusters are about $3$ fm, which indicates that the $ND$ state should be a molecular state with large separation. The formation of this structure is related to the weak interaction between two clusters; (3), for the coupling of two spatial structures, the system will automatically select a more favorable structure, because of its lowest energy. Meanwhile, the result indicates that the influence of channel coupling will pull the two clusters closer. This feature implies that the lowest state $\Sigma_c\pi$ has a relatively compact structure.
\begin{table}[h]
    \caption{\label{tab:table6}The RMS distances of the $qqq\bar{q}c$ pentaquark system with quantum numbers $I(J^P) = 0(\frac{1}{2}^-)$ (unit: fm).}
    \begin{ruledtabular}
    \begin{tabular}{ c c c c c }
        Structure & $r_{qq}$ & $r_{q\bar{q}}$ & $r_{qc}$ & $r_{c\bar{q}}$ \\
        \hline
        $[qqc][\bar{q}q]$ & $1.2$ & $1.1$ & $1.0$ & $1.2$ \\
        $[qqq][\bar{q}c]$ & $0.7$ & $2.9$ & $2.8$ & $0.6$ \\
        $coupling$ & $1.1$ & $1.0$ & $0.9$ & $1.0$ \\
    \end{tabular}
    \end{ruledtabular}
\end{table}

$\bm{I(J^P) = 0(\frac{3}{2}^-):}$ Table~\ref{tab:table7} lists the results of the single channel and channel coupling calculations. For the single channel calculation, we have found that the $\Sigma_c^* \pi$, $N D^*$ and $\Sigma_c \rho$ states can form to be bound among all possible physical channels, and the corrected energies of these three states are $2648$ MeV, $2943$ MeV and $3224$ MeV, respectively. For the channel coupling calculation, we obtain a lowest state of system, whose main component is the color singlet channel of $\Sigma_c^*\pi$, taking the percentage of $97.6\%$. This state with the energy of $2639$ MeV, is not far from the mass of $\Lambda_c(2625)$ with $J^P = \frac{3}{2}^-$. So for this charmed baryon, besides the description of three quark baryon in an orbital excitation, pentaquark component also becomes a possibility within the quark model calculation. Especially, we shift the energy of $N D^*$ state to $2943$ MeV, which is close to the mass of $\Lambda_c(2940)$. Our results support the existence of $\Lambda_c(2940)$ as a state composed by a nucleon and $D^*$ meson with $J^P = \frac{3}{2}^-$, as similar conclusions obtained in Ref~\cite{Zhao:2016zhf,He:2010zq,Ortega:2012cx,Zhang:2012jk}. Furthermore, another state $\Sigma_c \rho$ has disappeared in the channel coupling calculation, and the reason is that the channel coupling will push the higher state above the threshold.
\begin{table}[h]
    \caption{\label{tab:table7}The energies of the  $qqq\bar{q}c$ pentaquark system with quantum numbers $I(J^P) = 0(\frac{3}{2}^-)$ (unit: MeV).}
    \begin{ruledtabular}
    \begin{tabular}{ c c c c c c }
        Structure & Channel & $E$ & $E_{th}$(Theo.) & $E_B$ & $E'$ \\
        \hline
        $[qqc][\bar{q}q]$ & $(\Sigma_c^* \pi)^S$ & $2641$ & $2651$ & $-10$ & $2648$ \\
        {} & $(\Sigma_c^* \pi)^H$ & $3293$ & {} & {} & {} \\
        {} & $(\Lambda_c \omega)^S$ & $3044$ & $3043$ & $0$ & $3069$ \\
        {} & $(\Lambda_c \omega)^H$ & $3056$ & {} & {} & {} \\
        {} & $(\Sigma_c \rho)^S$ & $3258$ & $3264$ & $-6$ & $3224$ \\
        {} & $(\Sigma_c \rho)^H$ & $3332$ & {} & {} & {} \\
        {} & $(\Sigma_c^* \rho)^S$ & $3290$ & $3290$ & $0$ & $3293$ \\
        {} & $(\Sigma_c^* \rho)^H$ & $3271$ & {} & {} & {} \\
        $[qqq][\bar{q}c]$ & $(N D^*)^S$ & $2943$ & $2946$ & $-3$ & $2943$ \\
        {} & $(N D^*)^H$ & $3193$ & {} & {} & {} \\
        \hline
        $coupling$ & {} & $2633$ & $2651$ & $-18$ & $2639$ \\
        {} & \multicolumn{5}{c}{$(\Sigma_c^*\pi)^S$:$97.6\%$ ; $(ND^*)^S$:$1.2\%$ ; others:$1.2\%$} \\
    \end{tabular}
    \end{ruledtabular}
\end{table}

Different from the previous situation, the $ND^*$ state of $[qqq][\bar{q}c]$ structure can be bound in the single channel calculation. We analyze its energy components and present the result in Table~\ref{tab:table8}. Since the three identical quarks are restricted in the 3-quark cluster, kinetic energy of relative motion between two clusters provides less repulsion than the previous case, and contributions of confinement potential and one-gluon-exchange nearly all come from two subclusters, which shows that the benefit of confinement potential and one-gluon-exchange of forming a bound state can be neglected here. Nevertheless, both $\pi$ meson exchange and $\sigma$ meson exchange play an important role in binding a nucleon and $D^*$ meson together. As one can see in Table~\ref{tab:table9}, in which the RMS distances of the two structures and their coupling are calculated, similar results are obtained as before. For the $[qqq][\bar{q}c]$ structure, the distances between two hadrons are about twice as much as the $[qqc][\bar{q}q]$ structure. This motivates the possibility of finding molecular baryon-meson structure for $ND^*$ state. When coupling two structures, the system is stable in a structure with lower energy.
\begin{table}[h]
    \caption{\label{tab:table8}The energy components of $ND^*$ state (unit: MeV).}
    \begin{ruledtabular}
    \begin{tabular}{ c c c c c }
            {} & rest mass & kinetic & $V^C$ & $V^G$ \\
            \hline
            $ND^*$ & $3550.0$ & $1021.2$ & $-166.6$ & $-1097.0$ \\
            $N + D^*$ & $3550.0$ & $986.5$ & $-167.2$ & $-1096.4$ \\
            $\Delta_E$ & $0.0$ & $34.7$ & $0.6$ & $-0.6$ \\
            \hline
            {} & $V^{\pi}$ & $V^{K}$ & $V^{\eta}$ & $V^{\sigma}$ \\
            \hline
            $ND^*$ & $-332.2$ & $0.0$ & $62.8$ & $-95.1$ \\
            $N + D^*$ & $-319.5$ & $0.0$ & $63.6$ & $-71.2$ \\
            $\Delta_E$ & $-12.7$ & $0.0$ & $-0.8$ & $-23.9$ \\
    \end{tabular}
    \end{ruledtabular}
\end{table}
\begin{table}[h]
    \caption{\label{tab:table9}The RMS distances of $qqq\bar{q}c$ pentaquark system with quantum numbers $I(J^P) = 0(\frac{3}{2}^-)$ (unit: fm).}
    \begin{ruledtabular}
    \begin{tabular}{ c c c c c }
        Structure & $r_{qq}$ & $r_{q\bar{q}}$ & $r_{qc}$ & $r_{c\bar{q}}$ \\
        \hline
        $[qqc][\bar{q}q]$ & $1.4$ & $1.3$ & $1.1$ & $1.4$ \\
        $[qqq][\bar{q}c]$ & $0.7$ & $2.6$ & $2.5$ & $0.8$ \\
        $coupling$ & $1.4$ & $1.3$ & $1.1$ & $1.3$ \\
    \end{tabular}
    \end{ruledtabular}
\end{table}

$\bm{I(J^P) = 0(\frac{5}{2}^-):}$ In this case there are only two channels needed to be considered: the color singlet channel and the hidden color channel of $\Sigma_c^* \rho$. The results of the calculations are shown in Table~\ref{tab:table10}. When taking into account only the color singlet channel of $\Sigma_c^* \rho$, the calculation shows that $\Sigma_c^*$ baryon and $\rho$ meson can be attracted together to form a slightly bound state, with the binding energy of $-7$ MeV. When coupling to its hidden color channel, the energy of $\Sigma_c^* \rho$ decreases just $1$ MeV. This indicates that the contribution of the hidden color configuration to the bound state wave function is negligible, which only takes the percentage of $0.3\%$. As a consequence, we predict the existence of a high-spin state, $\Sigma_c^* \rho$, with a mass of $3285$ MeV in the iso-scalar sector. Because of the instability of $\rho$ meson, it is considered to be a wide pentaquark resonance. The RMS distances shown in Table~\ref{tab:table11} indicate that the configuration of $\Sigma_c^* \rho$ state with $J^P = \frac{5}{2}^-$ is different from the preliminary analysis of $\Sigma_c^{(*)} \pi$ state, quark distances are around 2 fm because of the weak binging energy. As mentioned above, the real distances among the sub-cluster should be less than $1$ fm, while the distance between two clusters is about 2.2 fm. We remark herein that this characteristic could point to have a state of molecular nature.
\begin{table}[h]
    \caption{\label{tab:table10}The energies of the  $qqq\bar{q}c$ pentaquark system with quantum numbers $I(J^P) = 0(\frac{5}{2}^-)$ (unit: MeV).}
    \begin{ruledtabular}
    \begin{tabular}{ c c c c c c }
        Structure & Channel & $E$ & $E_{th}$(Theo.) & $E_B$ & $E'$ \\
        \hline
        $[qqc][\bar{q}q]$ & $(\Sigma_c^* \rho)^S$ & $3283$ & $3290$ & $-7$ & $3286$ \\
        {} & $(\Sigma_c^* \rho)^H$ & $3388$ & {} & {} & {} \\
        \hline
        $coupling$ & {} & $3282$ & $3290$ & $-8$ & $3285$ \\
        {} & {} & \multicolumn{4}{c}{$(\Sigma_c^*\rho)^S$:$99.7\%$ ; $(\Sigma_c^*\rho)^H$:$0.3\%$} \\
    \end{tabular}
    \end{ruledtabular}
\end{table}
\begin{table}[h]
    \caption{\label{tab:table11}The RMS distances of the $qqq\bar{q}c$ pentaquark system with quantum numbers $I(J^P) = 0(\frac{5}{2}^-)$ (unit: fm).}
    \begin{ruledtabular}
    \begin{tabular}{ c c c c c }
        Structure & $r_{qq}$ & $r_{q\bar{q}}$ & $r_{qc}$ & $r_{c\bar{q}}$ \\
        \hline
        $[qqc][\bar{q}q]$ & $2.0$ & $2.0$ & $1.5$ & $2.2$ \\
    \end{tabular}
    \end{ruledtabular}
\end{table}

$\bm{I(J^P) = 1(\frac{1}{2}^-):}$ Table~\ref{tab:table12} indicates that there exist almost no bound states in all possible single channels except the weekly bound $ND^*$ state with the binding energy of $-1$ MeV. When doing the channel coupling calculation, the lowest state of the system, $\Lambda_c\pi$, has not been pushed down below its corresponding threshold, which indicates that $\Lambda_c\pi$ is a scattering state. Moreover, the $ND^*$ state is pushed above its threshold through the channel coupling calculation. Therefore, we find no bound state of the system in this quantum number sector.
Nevertheless, nearby the $ND$ threshold, we find an energy of $2773$ MeV. By analyzing its composition, we find that the color singlet channels of $ND$, $\Lambda_c\pi$ and $\Sigma_c\pi$ account for $67.6\%$, $18.3\%$ and $13.2\%$, respectively. This implies that there is a possibility to exist the resonance of systems with large $ND$ component. After correction, we obtain an energy of $2786$ MeV, which is close to the mass of iso-vector charmed baryon $\Sigma_c(2800)$. On the other side, $\Sigma_c(2800)$ was also investigated in Ref~\cite{JimenezTejero:2009vq,Dong:2010gu,Sakai:2020psu}, with the conclusion that it can be identified with a dynamically generated resonance with a dominant $ND$ configuration. The calculation of distances in Table~\ref{tab:table13} reveals the molecular nature of $[qqq][\bar{q}c]$ structure as before, but obviously large distances of $[qqc][\bar{q}q]$ structure indicate that the lowest state $\Lambda_c\pi$ can not form a bound state, and the coupling of two structures does not play a significant role.
\begin{table}[h]
    \caption{\label{tab:table12}The energies of the $qqq\bar{q}c$ pentaquark system with quantum numbers $I(J^P) = 1(\frac{1}{2}^-)$ (unit: MeV).}
    \begin{ruledtabular}
    \begin{tabular}{ c c c c c c }
        Structure & Channel & $E$ & $E_{th}$(Theo.) & $E_B$ & $E'$ \\
        \hline
        $[qqc][\bar{q}q]$ & $(\Lambda_c \pi)^S$ & $2428$ & $2426$ & $0$ & $2426$ \\
        {} & $(\Lambda_c \pi)^H$ & $3067$ & {} & {} & {} \\
        {} & $(\Sigma_c \pi)^S$ & $2625$ & $2625$ & $0$ & $2595$ \\
        {} & $(\Sigma_c \pi)^H$ & $3328$ & {} & {} & {} \\
        {} & $(\Sigma_c \eta)^S$ & $3130$ & $3129$ & $0$ & $3003$ \\
        {} & $(\Sigma_c \eta)^H$ & $3357$ & {} & {} & {} \\
        {} & $(\Lambda_c \rho)^S$ & $3065$ & $3065$ & $0$ & $3061$ \\
        {} & $(\Lambda_c \rho)^H$ & $3042$ & {} & {} & {} \\
        {} & $(\Sigma_c \rho)^S$ & $3264$ & $3264$ & $0$ & $3230$ \\
        {} & $(\Sigma_c \rho)^H$ & $3312$ & {} & {} & {} \\
        {} & $(\Sigma_c \omega)^S$ & $3242$ & $3242$ & $0$ & $3238$ \\
        {} & $(\Sigma_c \omega)^H$ & $3314$ & {} & {} & {} \\
        {} & $(\Sigma_c^* \rho)^S$ & $3291$ & $3290$ & $0$ & $3293$ \\
        {} & $(\Sigma_c^* \rho)^H$ & $3258$ & {} & {} & {} \\
        {} & $(\Sigma_c^* \omega)^S$ & $3268$ & $3268$ & $0$ & $3301$ \\
        {} & $(\Sigma_c^* \omega)^H$ & $3284$ & {} & {} & {} \\
        $[qqq][\bar{q}c]$ & $(N D)^S$ & $2782$ & $2782$ & $0$ & $2808$ \\
        {} & $(N D)^H$ & $3191$ & {} & {} & {} \\
        {} & $(N D^*)^S$ & $2945$ & $2946$ & $-1$ & $2945$ \\
        {} & $(N D^*)^H$ & $3174$ & {} & {} & {} \\
        {} & $(\Delta D^*)^S$ & $3261$ & $3260$ & $0$ & $3239$ \\
        {} & $(\Delta D^*)^H$ & $3559$ & {} & {} & {} \\
        \hline
        $coupling$ & {} & $2428$ & $2426$ & $0$ & $2426$ \\
        {} & {} & \multicolumn{4}{c}{$(\Lambda_c\pi)^S$:$100.0\%$} \\
        {} & {} & $2773$ & $2782$ & $-9$ & $2786$ \\
        {} & {} & \multicolumn{4}{c}{$(ND)^S$:$67.6\%$ ; $(\Lambda_c\pi)^S$:$18.3\%$ ;} \\
        {} & {} & \multicolumn{4}{c}{$(\Sigma_c\pi)^S$:$13.2\%$ ; others:$0.9\%$} \\
    \end{tabular}
    \end{ruledtabular}
\end{table}
\begin{table}[h]
    \caption{\label{tab:table13}The RMS distances of the $qqq\bar{q}c$ pentaquark system with quantum numbers $I(J^P) = 1(\frac{1}{2}^-)$ (unit: fm).}
    \begin{ruledtabular}
    \begin{tabular}{ c c c c c }
        Structure & $r_{qq}$ & $r_{q\bar{q}}$ & $r_{qc}$ & $r_{c\bar{q}}$ \\
        \hline
        $[qqc][\bar{q}q]$ & $4.3$ & $4.3$ & $3.1$ & $5.2$ \\
        $[qqq][\bar{q}c]$ & $0.7$ & $3.5$ & $3.5$ & $0.6$ \\
        $coupling$ & $4.3$ & $4.3$ & $3.1$ & $5.2$ \\
    \end{tabular}
    \end{ruledtabular}
\end{table}

$\bm{I(J^P) = 1(\frac{3}{2}^-):}$ We omit the numerical results herein because that no bound states are found among all possible physical channels, and the channel coupling does not play a crucial role in the formation of bound states.

$\bm{I(J^P) = 1(\frac{5}{2}^-):}$ As shown in Table~\ref{tab:table14}, there only exists the bound state in the $\Delta D^*$ configuration among all possible single channels, with the slight binding energy of $-1$ MeV. When coupling to other higher channels, the binding energy of $\Delta D^*$ state is deepened to $-8$ MeV. Besides, the component calculation indicates that the channel coupling plays an important role in forming a bound state in this case. Be similar to the case of $\Sigma_c^*\rho$ with $I(J^P) = 0(\frac{5}{2}^-)$, the $\Delta D^*$ appears as a wide pentaquark resonance because of the instability of $\Delta$ baryon in our present study. This result is consistent with the conclusion of Ref~\cite{Zhao:2016zhf,Carames:2012bd}, in which the $\Delta D^*$ with $I(J^P) = 1(\frac{5}{2}^-)$ was considered to be an attractive state. As we can see from the Table~\ref{tab:table15}, the pure $\Delta D^*$ state is a molecular state with large separation, however, the distances between two clusters are significantly pulled close because of the strong influence of structural coupling.
\begin{table}[h]
    \caption{\label{tab:table14}The energies of the $qqq\bar{q}c$ pentaquark system with quantum numbers $I(J^P) = 1(\frac{5}{2}^-)$ (unit: MeV).}
    \begin{ruledtabular}
    \begin{tabular}{ c c c c c c c }
        Structure & Channel & $E$ & $E_{th}$(Theo.) & $E_B$ & $E'$ \\
        \hline
        $[qqc][\bar{q}q]$ & $(\Sigma_c^* \rho)^S$ & $3290$ & $3290$ & $0$ & $3293$ \\
        {} & $(\Sigma_c^* \rho)^H$ & $3340$ & {} & {} & {} \\
        {} & $(\Sigma_c^* \omega)^S$ & $3269$ & $3268$ & $0$ & $3301$ \\
        {} & $(\Sigma_c^* \omega)^H$ & $3312$ & {} & {} & {} \\
        $[qqq][\bar{q}c]$ & $(\Delta D^*)^S$ & $3259$ & $3260$ & $-1$ & $3238$ \\
        {} & $(\Delta D^*)^H$ & $3564$ & {} & {} & {} \\
        \hline
        $coupling$ & {} & $3252$ & $3260$ & $-8$ & $3239$ \\
        {} & \multicolumn{5}{c}{$(\Delta D^*)^S$:$82.3\%$ ; $(\Sigma_c^*\omega)^S$:$11.3\%$ ; others:$6.4\%$} \\
    \end{tabular}
    \end{ruledtabular}
\end{table}
\begin{table}[h]
    \caption{\label{tab:table15}The RMS distances of the $qqq\bar{q}c$ pentaquark system with quantum numbers $I(J^P) = 1(\frac{5}{2}^-)$ (unit: fm).}
    \begin{ruledtabular}
    \begin{tabular}{ c c c c c }
        Structure & $r_{qq}$ & $r_{q\bar{q}}$ & $r_{qc}$ & $r_{c\bar{q}}$ \\
        \hline
        $[qqc][\bar{q}q]$ & $4.3$ & $4.3$ & $3.1$ & $5.2$ \\
        $[qqq][\bar{q}c]$ & $1.2$ & $3.4$ & $3.3$ & $0.9$ \\
        $coupling$ & $1.2$ & $1.9$ & $1.7$ & $1.0$ \\
    \end{tabular}
    \end{ruledtabular}
\end{table}

$\bm{I(J^P) = 2(\frac{1}{2}^-,\frac{3}{2}^-,\frac{5}{2}^-):}$ The numerical results and the detailed discussion is omitted when isospin $I=2$. The reason is that we find no bound states both in the single channel calculation and the channel coupling calculation. Moreover, we are not particularly interested in states with such higher isospin.

\begin{table}[h]
    \caption{\label{tab:table16}Possible pentaquarks of $qqq\bar{q}b$ system (unit: MeV).}
    \begin{ruledtabular}
    \begin{tabular}{ c c c c c c c }
        $I(J^P)$ & Main channel & $E$ & $E_{th}$(Theo.) & $E_B$ & $E'$ \\
        \hline
        $0(\frac{1}{2}^-)$ & $\Sigma_b \pi$ & $5934$ & $5964$ & $-30$ & $5922$ \\
        $0(\frac{3}{2}^-)$ & $\Sigma_b^* \pi$ & $5951$ & $5974$ & $-23$ & $5950$ \\
        $0(\frac{5}{2}^-)$ & $\Sigma_b^* \rho$ & $6604$ & $6613$ & $-9$ & $6598$ \\
        $1(\frac{5}{2}^-)$ & $\Delta B^*$ & $6567$ & $6579$ & $-12$ & $6545$ \\
    \end{tabular}
    \end{ruledtabular}
\end{table}
\begin{table}[h]
    \caption{\label{tab:table17}The RMS distances of the $qqq\bar{q}b$ pentaquark system (unit: fm).}
    \begin{ruledtabular}
    \begin{tabular}{ c c c c c c c }
        $I(J^P)$ & Main channel & $r_{qq}$ & $r_{q\bar{q}}$ & $r_{qb}$ & $r_{b\bar{q}}$ \\
        \hline
        $0(\frac{1}{2}^-)$ & $\Sigma_b \pi$ & $1.2$ & $1.1$ & $0.9$ & $1.1$ \\
        $0(\frac{3}{2}^-)$ & $\Sigma_b^* \pi$ & $1.3$ & $1.2$ & $1.0$ & $1.2$ \\
        $0(\frac{5}{2}^-)$ & $\Sigma_b^* \rho$ & $1.8$ & $1.8$ & $1.3$ & $1.9$ \\
        $1(\frac{5}{2}^-)$ & $\Delta B^*$ & $1.2$ & $1.6$ & $1.4$ & $0.9$ \\
    \end{tabular}
    \end{ruledtabular}
\end{table}
Because of the heavy flavor symmetry, we also extend the study to the bottom case of $qqq\bar{q}b$ system, possible pentaquarks of which are listed in Table~\ref{tab:table16}.  In iso-scalar sector, we have found $\Sigma_b \pi$, $\Sigma_b^* \pi$ and $\Sigma_b^* \rho$ states, while in the iso-vector sector, only found the $\Delta B^*$ state. Noting that  $\Sigma_b \pi$ and $\Sigma_b^* \pi$ could be the candidates of two odd parity bottomed baryons, $\Lambda_b(5912)$ and $\Lambda_b(5920)$, respectively. The RMS distances are listed in Table~\ref{tab:table17}. We consider these bottomed pentaquarks are more compact than those of the charm case.

In addition, we repeat the above calculations with another set of parameters to test the parameter sensitivity of the results. We find that although the binding energy has a small change, the qualitative conclusion of the existence of pentaquarks is independent of the parameters.

\section{\label{sec:level4}Summary}
In this work, we perform a dynamical calculation of $qqq\bar{q}Q$ ($q = u$ or $d$, $Q=c$ or $b$) five-quark system in the framework of the chiral quark model (ChQM) with Gaussian expansion method (GEM).

We first systematically study the system of $qqq\bar{q}c$, and then extend the calculation to the bottom case. Several charmed and bottomed pentaquarks with negative parity have been identified. Some excited charmed or bottomed baryons composed of $qqQ$ ($q = u$ or $d$, $Q=c$ or $b$), have a possibility to be pentaquark states except the explanation of three-quark baryons in an orbital excitation. Our results show that: (1),  $\Sigma_c\pi(IJ^P=0\frac{1}{2}^-)$ and $\Sigma_c^*\pi(IJ^P=0\frac{3}{2}^-)$ are possible resonances in our quark model calculation, and could be candidates of $\Lambda_c(2595)$ and $\Lambda_c(2625)$, respectively. Moreover, two high-spin pentaquark states, $\Sigma_c^*\rho(IJ^P=0\frac{5}{2}^-)$ and $\Delta D^*(IJ^P=1\frac{5}{2}^-)$, are found in the energy region of $3.2 \sim 3.3$ GeV; (2), With extension to the bottom case, $\Sigma_b\pi(IJ^P=0\frac{1}{2}^-)$, $\Sigma_b^*\pi(IJ^P=0\frac{3}{2}^-)$ could be candidates of $\Lambda_b(5912)$ and $\Lambda_b(5920)$, respectively. And $\Sigma_b^*\rho(IJ^P=0\frac{5}{2}^-)$ and $\Delta B^*(IJ^P=1\frac{5}{2}^-)$ are also found as resonances in the energy region of $6.5 \sim 6.6$ GeV; (3), The calculation of RMS distances reveals the compact nature for $\Sigma_c^{(*)}\pi$ and $\Sigma_b^{(*)}\pi$ states, and the molecular nature for $\Sigma_c^*\rho$, $\Sigma_b^*\rho$, $\Delta D^*$ and $\Delta B^*$ states.

Finally, we hope that more efforts could be devoted to investigating these excited baryons, and the pentaquarks predicted in this work are worth searching in future experiments. In addition, it is worthwhile to mention that the P-wave baryon will mix with the S-wave pentaquark. Therefore, study of baryons within the unquenched quark model, including the high Fock components, is our future work.

\begin{acknowledgments}
B.R.~He was supported in part by the National Natural Science Foundation of China (Grant No. 11705094), Natural Science Foundation of Jiangsu Province, China (Grant No. BK20171027), Natural Science Foundation of the Higher Education Institutions of Jiangsu Province, China (Grant No. 17KJB140011), and by the Research Start-up Funding (B.R.~He) of Nanjing Normal University.
And the work of J.L.~Ping was supported in part by the National Natural Science Foundation of China under Grants No. 11775118, and No. 11535005.
\end{acknowledgments}

\end{document}